# Efficient Data-Driven Production Scheduling in Pharmaceutical Manufacturing

Ioannis Balatsos[b], Athanasios Liakos[b], Panagiotis Karakostas[c], Tao Song[a], Vassilios Pantazopoulos[a], Christos Papalitsas[b]

[a]*Pfizer, Scientific Computing & HPC, Cambridge MA 02139, USA,*
[b]*Pfizer, Scientific Computing & HPC, Thessaloniki 55535, Greece,*
[c]*Department of Applied Informatics, School of Information Sciences, University of Macedonia,*
*156 Egnatia Str., Thessaloniki 54636, Greece,*

## Abstract

This paper develops a data-driven, constraint-based optimization framework for a complex industrial job shop scheduling problem variant in pharmaceutical manufacturing. The formulation captures fixed routings and designated machines, explicit resource calendars with weekends and planned maintenance, and campaign sequencing through sequence-dependent cleaning times derived from site tables. The model is implemented with an open source constraint solver and evaluated on deterministic snapshots from a solid oral dosage facility under three objective formulations: makespan, makespan plus total tardiness, and makespan plus average tardiness. On three industrial instances of increasing size (10, 30, and 84 jobs) the proposed schedules dominate reference plans that solve a simplified variant without the added site rules. Makespan reductions reach 88.1%, 77.6%, and 54.9% and total tardiness reductions reach 72.1%, 58.7%, and 18.2%, respectively. The composite objectives further decrease late job counts with negligible makespan change on the smaller instances and a modest increase on the largest instance. Optimality is proven on the small case, with relative gaps of 0.77% and 14.92% on the medium and large cases under a fixed time limit. The results show that a compact constraint programming formulation can deliver feasible, transparent schedules that respect site rules while improving adherence to due dates on real industrial data.

Modeling, Constraint Programming, Pharmaceuticals

## 1. Introduction

The pharmaceutical industry constitutes an important sector of the chemical process industry, playing a pivotal role in both global healthcare and the economy [1]. This industry is characterized by high regulatory standards, complex production processes, and stringent quality control requirements [2]. The increased need to develop and produce new safe and effective drugs requires significant investments in research and development operations, while economic growth must be achieved simultaneously [3]. Additionally, given the increasing demand for efficiency and compliance with Good Manufacturing Practices for biological products of the World Health Organization [4, 5], pharmaceutical manufacturing faces significant challenges in optimizing production schedules [6]. Therefore, efficient scheduling in pharmaceutical production is essential to reduce lead times, minimize costs, and ensure timely drug availability while maintaining high product quality and adherence to strict regulations [7, 8].

Pharmaceutical manufacturing involves a diverse set of processes, including raw material preparation, granulation, blending, compression, filtration, coating, packaging, and quality control testing. These processes often require shared resources, such as specialized machinery, cleanrooms, and skilled personnel, making scheduling highly complex [9]. Additionally, pharmaceutical production is constrained by factors such as batch processing requirements, machine cleaning and setup times, campaign scheduling for different drug formulations, and regulatory compliance constraints [10]. This complexity makes traditional scheduling approaches inefficient, necessitating the use of advanced optimization techniques .

Combinatorial optimization plays a crucial role in addressing the challenges of production scheduling in pharmaceutical manufacturing [11]. Among the various combinatorial optimization problems, the Job Shop Scheduling Problem (JSSP) is particularly relevant due to its applicability in environments where multiple jobs must be processed on a set of machines with specific operational constraints. JSSP is known to be NP-hard, making it computationally challenging to solve optimally for large-scale instances. In industrial practice, such as pharmaceutical manufacturing, job-shop formulations must model sequence-dependent setup times due to cleaning, machine-availability calendars [12], and campaign scheduling for enhanced operational efficiency.

To tackle these challenges, researchers and industry practitioners employ mathematical programming techniques such as Mixed-Integer Linear Programming, Constraint Programming [13, 14], and heuristic and metaheuristic approaches. This paper develops a data-driven, constraint-based optimization framework for a rich job shop scheduling problem in pharmaceutical manufacturing. The formulation captures fixed routings and designated machines, explicit machine-availability calendars with weekends and planned maintenance, and sequence-dependent cleaning times that induce campaign sequencing. Inputs are deterministic snapshots extracted from manufacturing systems. The model is implemented with an open-source solver (OR-Tools) and evaluated on real site data under three objective formulations: makespan, makespan plus total tardiness, and makespan plus average tardiness. The results show that a compact constraint programming formulation can deliver feasible, transparent schedules that respect site rules while improving adherence to due dates.

The remainder of the paper is organized as follows. Section 2 states the industrial scheduling setting and the scope of the data snapshots. Section 3 presents the constraint-based model, including time discretization, decision variables, constraints, and the objective formulations. Section 4 reports computational studies on real site data, covering model validation, scaling across instance sizes, and the impact of the objective function. Section 5 concludes with key insights and directions for future research.

## 2. Problem statement

Scheduling is a key operational activity in manufacturing, directly influencing both the effectiveness and efficiency of production systems. Optimizing such complex decision-making processes requires robust methodologies from Operations Research, particularly combinatorial optimization. One of the most extensively studied problems in this context is the Job Shop Scheduling Problem (JSSP) [15], which models the sequencing of operations on multiple machines to optimize performance measures such as makespan, total completion time, or tardiness [16]. The classical JSSP assumes that each job follows a predefined processing sequence across a fixed set of machines, with no flexibility in machine assignment [17].

Despite advancements in production scheduling models, real facilities require formulations that reflect site-specific policies and calendars beyond the basic job shop abstraction. At a site of a global pharmaceutical manufacturer producing solid oral dosage forms, scheduling must respect fixed routings per job, eligible machines per operation, machine calendars with weekends and planned maintenance, and sequence-dependent cleaning between consecutive operations. The inputs used in this study are deterministic snapshots of released orders and resource calendars. Demand and supply variability are handled upstream and are outside the scope of this work.

The industrial data therefore, describes a flexible job shop with eligible machines defined for many operations. Machine choice is resolved in a preprocessing step through a load spreading heuristic that assigns each operation to one eligible machine in order to control per-machine task counts and to reduce sequencing complexity. After preprocessing, the resulting instance is a rich job shop with fixed machine assignments, sequence-dependent cleaning derived from site rules based on product family, active ingredient strength, and operation family, and resource calendars with global non-working days and machine-specific maintenance. The decision is to set start and finish times for all operations and to select a feasible order on each machine while respecting calendars and cleaning requirements, and while evaluating the performance criteria stated below.

Within this setting, the manufacturing team must generate large-scale schedules within practical time limits while enforcing the following operational constraints:

- precedence within each job that mandates a strict, predefined order of operations

- machine capacity with at most one operation processed at a time

- campaign sequencing on each machine with sequence-dependent cleaning times between consecutive operations

- maintenance windows that reflect the actual availability of specific machines

- calendar exclusions for non-working days such as weekends and scheduled vacations

Incorporating these elements into a unified optimization framework is essential for improving production efficiency and meeting the high operational standards of the global pharmaceutical manufacturer under study

In this study the baseline performance criterion is the makespan of the production plan. Makespan is a standard efficiency metric that captures the total duration of the plan. To reflect adherence to due dates we also analyze one alternative criterion defined on the same feasible set, namely the sum of the makespan and the total tardiness across all jobs. Tardiness is the positive deviation of a job completion time from its due date, and earliness is not rewarded. All terms use the same time units, so the sum is directly interpretable without ad hoc weights.

*2.1. Industrial manufacturing scheduling context*

The study is located at the site of a global pharmaceutical manufacturer that produces solid oral dosage forms. Over time the site has progressed from manual spreadsheet practices and physical visual boards to a commercial scheduling platform. This transition increased baseline discipline, yet the platform provides limited control over the full set of shop floor rules and limited transparency for analytical review. These limitations motivate a formal model that is faithful to plant policies and that can be audited by operations, planning, and quality.

Scheduling at the site operates on deterministic snapshots of released orders and resource calendars. Jobs follow fixed routings and many operations have eligible machines recorded in the manufacturing systems. Machine choice is resolved before optimization by a preprocessing heuristic that assigns each operation to one eligible machine so that per-machine load is controlled and sequencing complexity is reduced. Non-working days and planned maintenance are known in advance and must be respected. Consecutive operations on the same machine require sequence-dependent cleaning times that follow site rules and tables. Within this setting, the scheduling task is to produce large-scale timetables within practical time limits while maintaining full traceability to the underlying data and procedures.

To meet these needs, the site is pursuing a constraint-based scheduling capability developed in collaboration with the Scientific Computing and High-Performance Computing team of the manufacturer under study. The approach encodes site-specific calendars and data-driven cleaning times in a single rigorous formulation that connects directly to manufacturing data sources. Execution on high-performance infrastructure enables turnaround times that align with operational decision cycles and supports frequent updates when new snapshots become available.

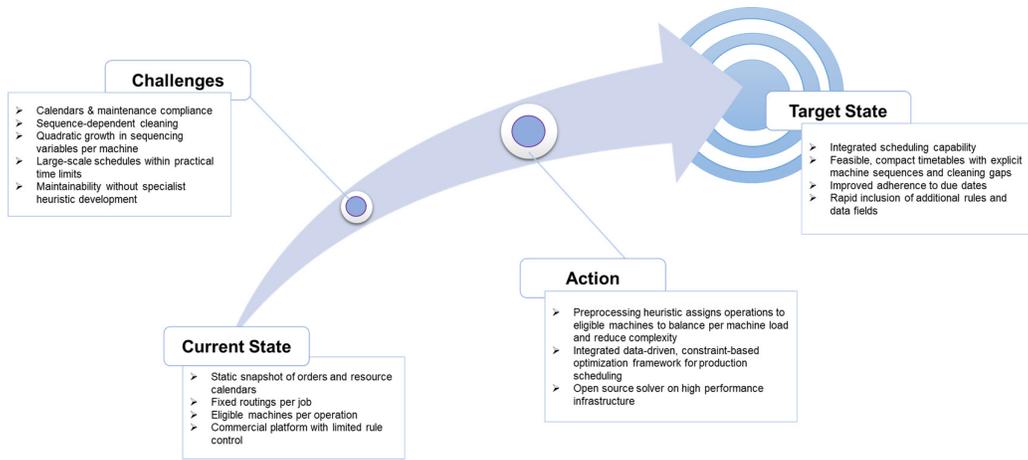

Figure 1: Synopsis of the work under consideration

The intended outcome is a reliable and transparent scheduling process that delivers compact and feasible production plans, improves adherence to due dates, and makes explicit the trade-offs induced by cleaning requirements and calendar restrictions. The capability is designed to support rapid re-optimization and to provide schedules that can be reviewed and approved within the existing governance of the site.

As a concise summary of the industrial context and the modelling roadmap, Figure 1 illustrates the current state, key challenges, proposed approach, and target state.

## 3. Constraint programming model formulation

This section presents a self-contained constraint based formulation for the scheduling setting described above. Jobs consist of ordered tasks (operations). A preprocessing step resolves machine choice among eligible alternatives through a data-driven load-spreading heuristic, after which each task is assigned to a designated machine. Machines have non-working periods, including weekends, holidays, and planned maintenance. Consecutive tasks processed on the same machine require sequence-dependent cleaning or setup time that depends on the ordered pair of tasks through product family, active ingredient strength, and operation family. The formulation is expressed in discrete time.

*3.1. Time discretisation and horizon*

All time quantities are represented as integers using a scaling factor $s = 10^L$, where $L$ is the maximum number of decimal digits observed in the input data. Let $H \in \mathbb{Z}_{>0}$ denote an upper bound on the planning horizon, computed as the sum of processing times with a conservative allowance for sequence-dependent cleaning. The time origin $t = 0$ corresponds to the start timestamp of the scheduling, and all due dates and calendar intervals are assigned and scaled to this origin.

*3.2. Sets and indices*

- $J = \{1, \ldots, n\}$: set of jobs (index $j$).

- For each $j$, $T_j = \{1, \ldots, |T_j|\}$: ordered set of tasks of job $j$ (index $t$).

- $M$: set of machines (index $m$).

- For each task $(j, t)$, the execution machine is fixed after preprocessing and is denoted $m(j, t) \in M$.

- $W$: set of global non-working intervals applied to all machines (for example, weekends and holidays).

- For each $m$, $R_m$: set of planned maintenance intervals specific to machine $m$.

*3.3. Parameters*

- $p_{j,t} \in \mathbb{Z}_{>0}$: processing time of task $(j, t)$ (already scaled by $s$).

- $H \in \mathbb{Z}_{>0}$: time horizon upper bound.

- For each $w \in W$, a fixed interval $[a_w, b_w) \subset [0, H]$ represents a global non-working period.

- For each $m \in M$ and $r \in R_m$, a fixed interval $[a_{m,r}, b_{m,r}) \subset [0, H]$ represents the maintenance of the machine $m$.

- $d_j \in \mathbb{Z}$: due date of the job $j$ measured on the same time scale; every job has a due date.

- $\kappa_{\alpha \to \beta} \in \mathbb{Z}_{\geq 0}$: sequence dependent cleaning or setup time when task $\beta$ immediately follows task $\alpha$ on the same machine. The value is computed by a data-driven function $\text{CleanCost}(\alpha, \beta)$ that uses the attributes noted above, with default categories when a specific entry is not available.

*3.4. Decision variables*

*Task–level variables..* For each task $(j, t)$ introduce start and end times and a non–non-preemptive interval:

$$\begin{aligned} S_{j,t} &\in [0, H], &&\text{start time of task } (j, t), \\ E_{j,t} &\in [0, H], &&\text{end time of task } (j, t), \\ I_{j,t} &= [S_{j,t}, E_{j,t}), &&\text{non–preemptive interval}, \\ E_{j,t} &= S_{j,t} + p_{j,t}. \end{aligned}$$

*Job– and plan–level variables..* For each job $j$,

$$C_j := E_{j,|T_j|}, \qquad T_j \geq 0,$$

and the makespan satisfies

$$C_{\max} \in [0, H], \qquad C_{\max} \geq C_j \;\; \forall j \in J.$$

*Machine–level sequencing variables..* For each machine $m$ and each ordered pair of distinct tasks $\alpha \neq \beta$ with $m(\alpha) = m(\beta) = m$, introduce

$$z_{\alpha \to \beta} \in \{0, 1\},$$

which indicates that $\beta$ is scheduled immediately after $\alpha$ on machine $m$. Dummy start and end nodes are included in the usual way.

*3.5. Constraints*

(a) Intra job precedence.

$$S_{j,t+1} \geq E_{j,t}, \qquad \forall j \in J, \; t = 1, \ldots, |T_j| - 1. \tag{1}$$

*(b) Machine capacity and calendars.* Tasks assigned to the same machine cannot overlap in time and must respect global non-working periods and machine maintenance windows. This is enforced with a `NoOverlap` constraint:

For each machine $m$, define

$$\mathcal{T}_m := \{ I_{j,t} \ : \ m(j,t) = m \},$$
$$\mathcal{W}_m := \{ I^{\text{idle}}_{w,m} := [a_w, b_w) \ : \ w \in W \},$$
$$\mathcal{R}_m := \{ I^{\text{mnt}}_{m,r} := [a_{m,r}, b_{m,r}) \ : \ r \in R_m \}.$$

Non-overlap with calendars and maintenance is then enforced by the numbered constraint

$$\text{NoOverlap}(\mathcal{T}_m \cup \mathcal{W}_m \cup \mathcal{R}_m), \qquad \forall m \in M. \tag{2}$$

Here $I^{\text{idle}}_{w,m}$ denotes the replication of each global non-working interval on machine $m$.

*(c) Campaign sequencing with sequence dependent setups.* For each machine $m$, the chosen order of its tasks forms a single directed circuit:

$$\text{Circuit}(\{z_{\alpha \to \beta}\}), \qquad \text{on all tasks with } m(\alpha) = m(\beta) = m. \tag{3}$$

Sequence-dependent cleaning or setup times are imposed through reified precedences:

$$z_{\alpha \to \beta} = 1 \ \Rightarrow \ S_\beta \geq E_\alpha + \kappa_{\alpha \to \beta}, \qquad \forall \alpha \neq \beta : \ m(\alpha) = m(\beta). \tag{4}$$

The combination of (2)–(3)–(4) ensures non-overlap on each machine, a unique processing order, and proper inclusion of sequence-dependent cleaning without big $M$ constants. The function $\text{CleanCost}(\alpha, \beta)$ is machine-dependent and asymmetric in general, since it is defined on ordered pairs of tasks.

*(d) Makespan.*
$$C_{\max} \geq C_j, \qquad \forall j \in J. \tag{5}$$

*(e) Tardiness.*
$$T_j \geq C_j - d_j, \qquad T_j \geq 0, \qquad \forall j \in J. \tag{6}$$

*3.6. Objective functions*

Three objectives are considered on the same feasible set. The primary criterion is the minimization of makespan, which measures the overall length of the schedule. Two alternative criteria retain the makespan term and add a tardiness component to capture due-date performance, namely, total tardiness and average tardiness. Keeping the makespan term preserves incentives for schedule compaction and guards against solutions that satisfy due dates by unnecessarily dilating the plan.

*(1) Makespan .*

$$\min \ C_{\max}. \tag{7}$$

*(2) Makespan plus total tardiness .*

$$\min \ C_{\max} \ + \ \sum_{j \in J} T_j. \tag{8}$$

*(3) Makespan plus average tardiness .*

$$\min \ C_{\max} \ + \ \frac{1}{|J|} \sum_{j \in J} T_j. \tag{9}$$

All terms use the same time scale, and earliness is not rewarded. We adopt simple additive forms without weights to maintain transparent interpretation and avoid ad hoc parameter tuning.

*3.7. Data-driven cleaning and setup times*

The quantity $\kappa_{\alpha \to \beta}$ is obtained from the function $\text{CleanCost}(\alpha, \beta)$ that (i) identifies the operation family from the machine, (ii) queries site tables parameterised by product family and active ingredient strength, and (iii) falls back to default categories when a specific pair is not present. All cleaning times are scaled by $s$ and treated as nonnegative integers.

## 4. Case studies & results

*4.1. Experimental Setup*

All experiments were executed on an enterprise-grade compute node within a secured industrial computing environment made available to the project team. The host includes a Dell PowerEdge R760xa chassis provisioned with four NVIDIA A100 GPUs. However, the solver and model are CPU-bound, and no computation was offloaded to GPUs. Jobs were launched via the site scheduler (Slurm 24.05.4) as single-node runs under identical policy settings. The implementation is in Python 3.10.6 using OR-Tools 9.14 for constraint programming. Time is discretized via an integer scaling factor determined by the maximum number of decimal digits in the input data, and the same codebase and parameters are used across all experiments.

Time is discretized via an integer scaling factor determined by the maximum number of decimal digits in the input data, and the same codebase and parameters are used across all experiments. Unless otherwise specified, the CPU time limit is 7,200 seconds for the medium and large instances, while the small snapshot proves optimality within seconds.

The evaluation uses three industrial snapshots with 10, 30, and 84 jobs. Each snapshot is a deterministic extract of released orders and resource calendars (weekends, site holidays, and machine-specific maintenance). A data-driven preprocessing step assigns each operation to an eligible designated machine. Sequence-dependent cleaning times are derived from site tables parameterised by product family, active ingredient strength, and operation family. Every job has a due date. On this common feasible set we compare three objective formulations—makespan (primary), makespan plus total tardiness, and makespan plus average tardiness—in order to quantify the trade-off between schedule compactness and due-date adherence without confounding factors. For context, we also report the metrics of the site planning system in the same snapshots when available; these reference plans solve a simplified variant that omits some of the shop floor rules enforced here, so comparisons are indicative rather than strictly like-for-like. We report makespan (days), total/average/maximum tardiness (days), count of late jobs, solver wall-clock time (seconds), the objective value and best bound (model time units), and the relative optimality gap at termination

*4.2. Model validation*

The formulation is validated on a small industrial snapshot with 10 jobs from a solid oral dosage line. The instance carries the full constraint structure used at scale, including ordered operations within each job, designated machines after preprocessing, sequence-dependent cleaning between consecutive operations executed on the same machine, global non-working periods such as weekends and site holidays, machine-specific maintenance windows, and due dates for all jobs. Validation is carried out under the primary objective of makespan. Because the feasible region does not depend on the objective, the same checks apply to the alternative criteria.

An automated post-processing routine verifies that all precedence relations hold, that no two operations overlap on the same machine, that no operation is scheduled within global idle intervals or maintenance windows, and that the idle gap between consecutive operations on the same machine is at least the prescribed sequence-dependent cleaning time for the ordered pair. Tardiness is recomputed per job from due dates and completion times to confirm consistency with reported aggregates. Visual inspection of machine-level Gantt and campaign views corroborates the absence of capacity conflicts and the explicit presence of cleaning gaps.

All checks pass with no violations. The solver establishes optimality for this instance, and the resulting schedule is compact and compliant with site rules. Table 1 reports metrics for the validated schedule and for a reference schedule on the same problem instance.

Table 1: Validation metrics for the small industrial instance with 10 jobs under the makespan objective.

|  | Makespan | Total tard. | Avg. tard. | Max tard. [d] | Late jobs | Runtime [s] |
| --- | --- | --- | --- | --- | --- | --- |
| Validated schedule | 3.67 | 54.19 | 5.42 | 53.24 | 3 | 35 |
| Reference schedule | 30.89 | 194.48 | 19.45 | 68.64 | 8 | – |

Figure 2 shows the machine-level Gantt for the 10-job instance, illustrating the final sequencing and the compactness of the plan.

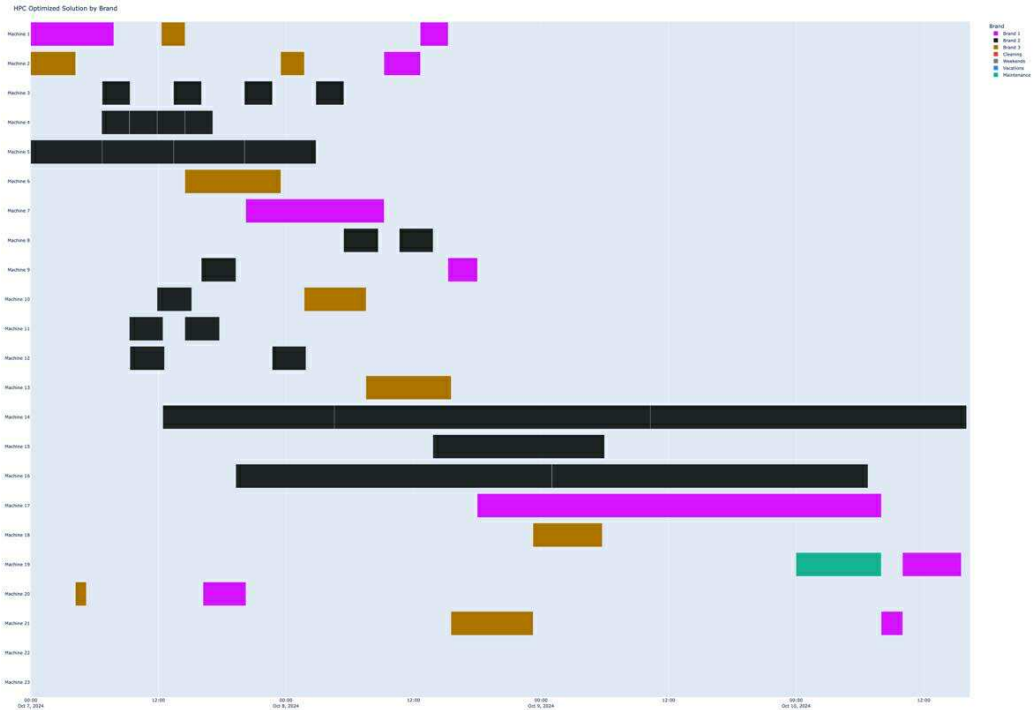

Figure 2: The Gantt-chart of the optimal production plan for the small-sized instance

These results indicate that the proposed formulation produces an optimal plan for the small-scale industrial case under consideration, while enforcing calendars, maintenance, and sequence-dependent cleaning by construction. The same automated checks were applied to all schedules reported in the present Section, and no feasibility violations were detected.

### 4.3. Computational scaling on site data snapshots

To assess computational scalability, the proposed model is evaluated on three real data snapshots of increasing size with 10, 30, and 84 jobs. All instances are built with the same preprocessing and calendar rules, and all runs use identical solver settings. The analysis focuses on the makespan objective, which aligns with the current operational target at the site.

Table 2 reports the quality of the schedule for the proposed model and for the reference plans, computed on the same snapshots. The proposed schedules exhibit systematically smaller makespans at all sizes. This observation is particularly significant because the reference plans solve a simplified variant that omits several site rules, whereas the proposed formulation enforces them. Achieving lower makespans within a stricter feasible region indicates more effective use of capacity and sequencing. Tardiness measures are also reduced while the model enforces sequence-dependent cleaning and machine calendars by construction. The reductions are substantial on the two smaller snapshots and remain positive on the largest. Late-job counts drop from 8 to 3 jobs at 10 jobs and from 23 to 11 jobs at 30 jobs, with a modest improvement from 50 to 48 at 84 jobs.

Table 2: Schedule quality across site data snapshots under the makespan objective

| Instance | Makespan [d] | Total tard. [d] | Avg. tard. [d] | Max tard. [d] | Late jobs | Plan |
|---|---|---|---|---|---|---|
| 10 jobs | 3.67 | 54.19 | 5.42 | 53.24 | 3 | Proposed |
| 10 jobs | 30.89 | 194.48 | 19.45 | 68.64 | 8 | Reference |
| 30 jobs | 8.98 | 192.60 | 6.42 | 59.94 | 11 | Proposed |
| 30 jobs | 40.07 | 466.25 | 15.54 | 68.64 | 23 | Reference |
| 84 jobs | 18.09 | 1717.92 | 20.45 | 176.11 | 48 | Proposed |
| 84 jobs | 40.07 | 2100.31 | 25.00 | 190.29 | 50 | Reference |

Figure 3 aggregates the performance gains as percentage reductions relative to the reference. Makespan falls by 88.1%, 77.6%, and 54.9% on the 10, 30, and 84 job snapshots, respectively. Total tardiness drops by 72.1% and 58.7% on the two smaller snapshots and by 18.2% on the largest. The diminishing marginal improvement with instance size is consistent with a fixed time limit and the increased sequencing complexity on heavily loaded machines. Nevertheless, the gains remain substantial across all sizes.

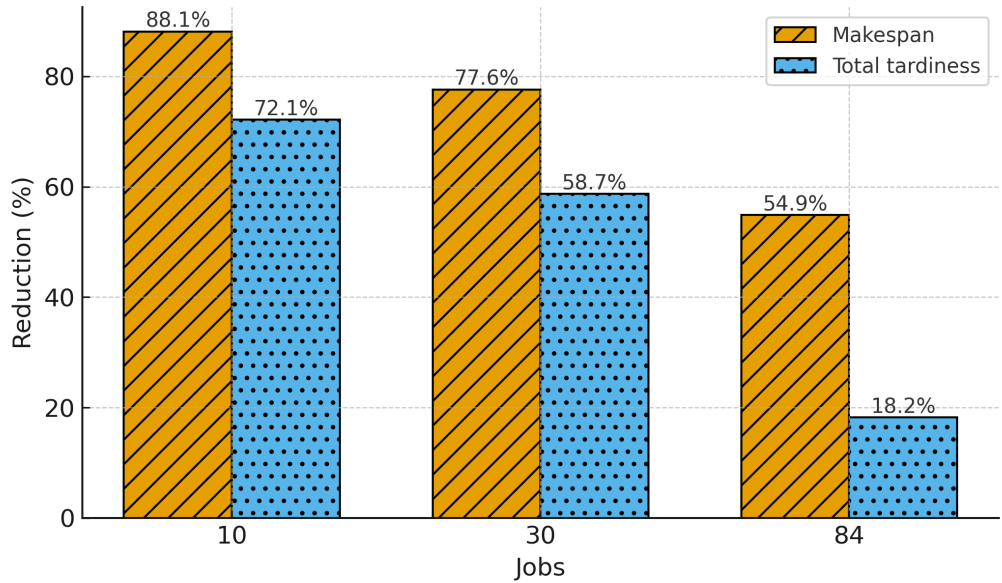

Figure 3: Relative improvement over the reference on the same snapshots

Table 3 summarises model size and solver outcomes. The count of campaign-sequencing literals increases from 236 at 10 jobs to 2,225 at 30 jobs and 7,850 at 84 jobs, which follows the quadratic growth in per-machine ordering choices. Optimality is certified on the 10-job snapshot. The medium snapshot stops at the time limit with a relative optimality gap of 0.77%. The large snapshot stops at the time limit with a gap of 14.92%. These gaps are consistent with the growth in the search space.

Table 3: Model size and solver outcomes under the makespan objective.

| Instance | Seq. literals | Tasks/machine (min–avg–max) | Runtime [s] | Status | Best bound | Gap [%] |
| --- | --- | --- | --- | --- | --- | --- |
| 10 jobs | 236 | 1–2.78–8 | 35 | Optimal | 8802 | 0.00 |
| 30 jobs | 2225 | 1–6.61–25 | 7200 | Feasible | 21386 | 0.77 |
| 84 jobs | 7850 | 1–10.59–45 | 7200 | Feasible | 36935 | 14.92 |

To clarify, sequencing literals are the per-machine arc variables used in campaign ordering.

Figure 4 plots the relative optimality gap of the solver against the number of literals of sequencing. The monotone pattern links proof quality directly to model size: as the number of campaign arcs grows, more reified precedences become candidates, bound tightening slows, and the remaining gap at a fixed time limit increases. This structural relationship explains why optimality is easy at 10 jobs, nearly proven at 30 jobs, and harder at 84 jobs.

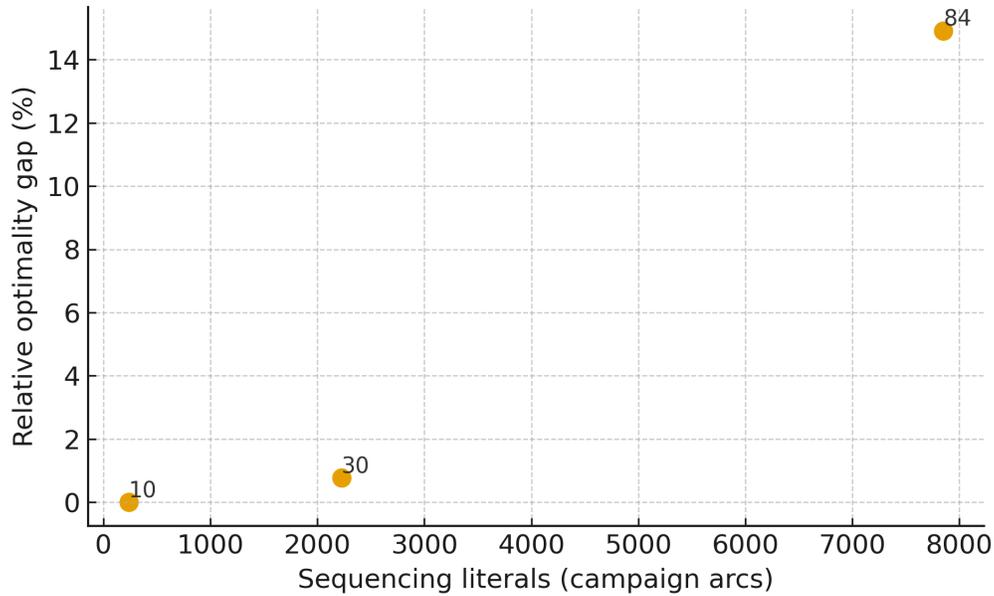

Figure 4: Relative optimality gap versus sequencing literals

Figure 5 shows the min–mean–max tasks per machine across snapshots. The average load rises from 2.78 to 6.61 and 10.59 tasks per machine, while the maximum rises from 8 to 25 and 45. Higher per-machine load implies more candidate successors per task, increasing the number of arcs $n_m(n_m-1)$ and the associated reified setup constraints. This load-driven expansion of the campaign search space aligns with the observed increase in optimality gaps and the gradual attenuation of percentage improvements as problem size grows.

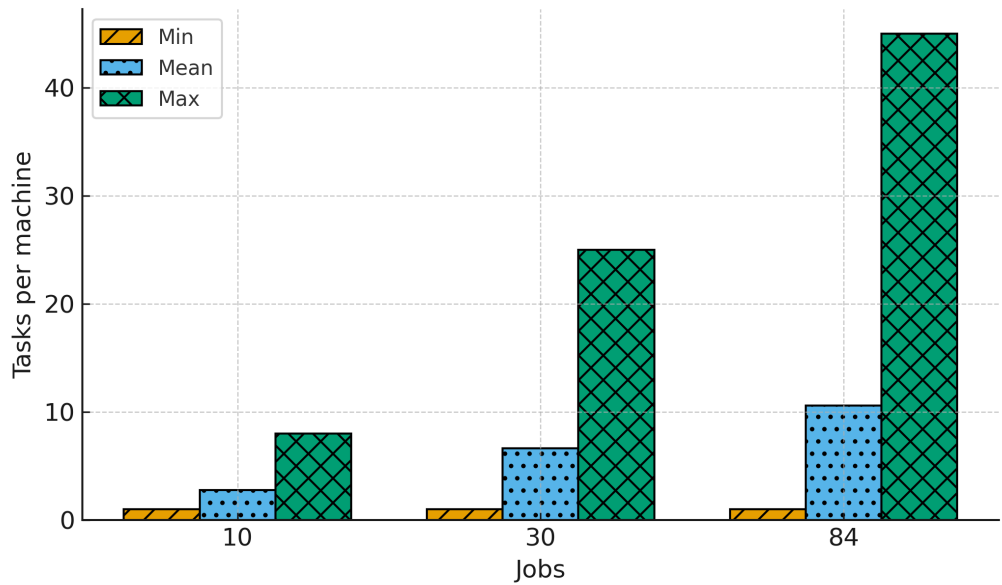

Figure 5: Per-machine task load across snapshots

In aggregate, the results indicate favourable scaling under site rules. The model delivers compact plans with sizable reductions in makespan and tardiness relative to the reference across all sizes. As instance size grows, the number of binary arc variables for campaign sequencing and the per-machine loads increase, which expands the ordering space and makes optimality certification harder within a fixed time budget. However, the quality of the solution remains strong, and all schedules comply with sequence-dependent cleaning requirements and site calendars.

## 4.4. Impact of objective function formulation

This subsection provides the analysis of how alternative objective functions affect the quality of the production schedule, under an identical feasible set. We examine three objective cases, one considering a pure makespan objective function, one combining makespan with average tardiness, and a combination between makespan and total tardiness. All other aspects of the experimental setup remain fixed, so observed differences reflect only the choice of objective. The analysis clarifies the trade-off between plan compactness and due-date adherence and indicates which formulation offers the best balance at each problem size.

Table 4 reports the main schedule metrics. On the small snapshot the composite objectives reduce late jobs from 3 to 2 with essentially unchanged makespan. On the medium snapshot they reduce total tardiness by 7.4%–9.5% and late jobs from 11 to 7, again with no makespan increase. On the large snapshot they reduce total tardiness by 8.6%–9.3% and late jobs by 23% (from 48 to 37) at a modest makespan increase of 2.3%–3.2%. Figure 6 visualizes total tardiness in days across objectives and instance sizes, and Figure 7 summarizes late-job counts.

Table 4: Schedule metrics by instance size and objective.

| Jobs | Objective | Makespan [d] | Total tard. [d] | Avg. tard. [d] | Max tard. [d] | Late jobs | Runtime [s] |
|---|---|---|---|---|---|---|---|
| 10 | Makespan | 3.67 | 54.19 | 5.42 | 53.24 | 3 | 35 |
| 10 | Makespan + Avg tard. | 3.67 | 53.91 | 5.39 | 53.24 | 2 | 34 |
| 10 | Makespan + Total tard. | 3.79 | 53.61 | 5.36 | 52.94 | 2 | 33 |
| 30 | Makespan | 8.98 | 192.60 | 6.42 | 59.94 | 11 | 7200 |
| 30 | Makespan + Avg tard. | 8.98 | 174.35 | 5.81 | 55.94 | 7 | 7200 |
| 30 | Makespan + Total tard. | 8.98 | 178.44 | 5.95 | 56.00 | 7 | 7200 |
| 84 | Makespan | 18.09 | 1717.92 | 20.45 | 176.11 | 48 | 7200 |
| 84 | Makespan + Avg tard. | 18.67 | 1569.92 | 18.69 | 175.60 | 37 | 7200 |
| 84 | Makespan + Total tard. | 18.50 | 1559.08 | 18.56 | 174.74 | 37 | 7200 |

Table 5 reports objective values, best bounds, and gaps. As instance size increases and as the objective aggregates more terms, optimality certificates become harder under the fixed wall-clock limit, but the solutions remain of high quality.

Table 5: Solver bounds and optimality gaps (objective values are in model time units).

| Jobs | Objective | Objective value | Best bound | Gap [%] | Status |
|---|---|---|---|---|---|
| 10 | Makespan | 8,802 | 8,802 | 0.00 | Optimal |
| 10 | Makespan + Avg tard. | 11,928 | 11,928 | 0.00 | Optimal |
| 10 | Makespan + Total tard. | 40,007 | 40,007 | 0.00 | Optimal |
| 30 | Makespan | 21,551 | 21,386 | 0.77 | Feasible |
| 30 | Makespan + Avg tard. | 25,455 | 23,656 | 7.07 | Feasible |
| 30 | Makespan + Total tard. | 139,862 | 95,018 | 32.06 | Feasible |
| 84 | Makespan | 43,411 | 36,935 | 14.92 | Feasible |
| 84 | Makespan + Avg tard. | 76,589 | 62,389 | 18.54 | Feasible |
| 84 | Makespan + Total tard. | 2,720,555 | 2,171,893 | 20.17 | Feasible |

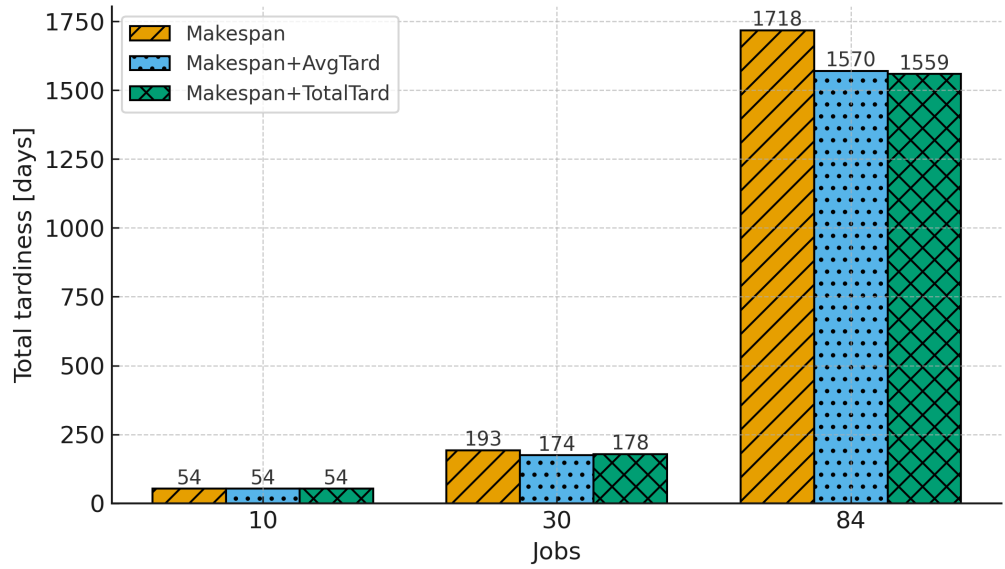

Figure 6: Total tardiness in days across objectives and instance sizes (hatch patterns aid grayscale print).

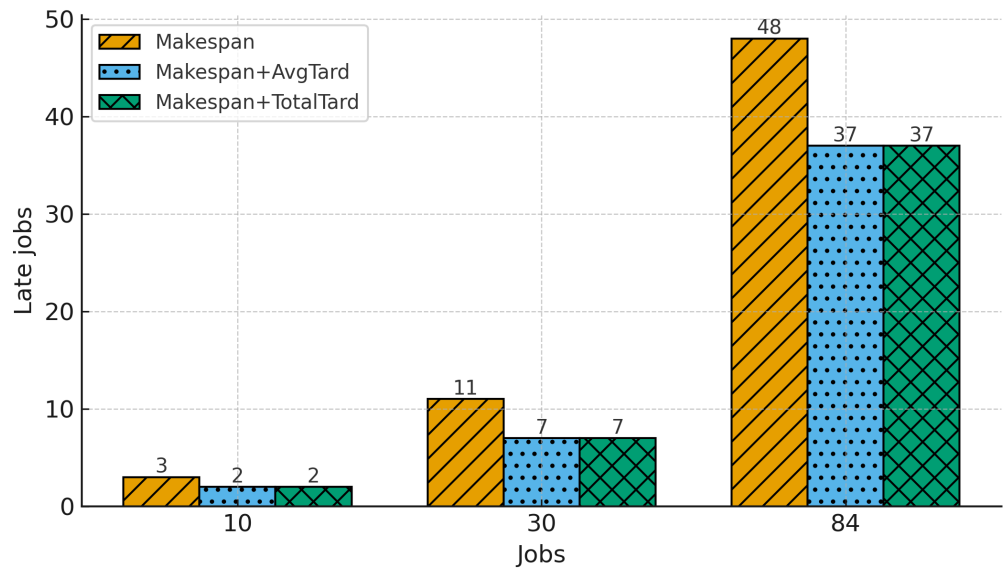

Figure 7: Late jobs across objectives and instance sizes.

Two systematic effects are observed across instance sizes. When machine loads are moderate (10 and 30 jobs), calendar exclusions and sequence-dependent cleaning leave exploitable slack. The composite objectives reallocate this slack to critical operations on due date, which reduces total tardiness and the number of late jobs without an increase in the makepan. When loads are high (84 jobs), the same reallocation requires extending campaign blocks or accepting longer idle gaps, so the makespan rises modestly while the tardiness still falls. The relative behavior of the two composite formulations is consistent with their emphasis, as the average-tardiness variant tends to spread improvements across many jobs and is effective at lowering the late-job count, whereas the total-tardiness variant prioritizes cutting aggregate delay and achieves the smallest total tardiness on the largest snapshot. Finally, the larger optimality gaps for the composite formulations at fixed CPU time are expected, since the richer objective introduces additional trade-offs that slow bound progress; this does not alter the qualitative improvements in due-date adherence.

## 5. Conclusion

This work presented a data-driven, constraint-based formulation for a complex job-shop scheduling problem that arises in pharmaceutical manufacturing. The model encodes fixed routings and designated machines, machine calendars with weekends and planned maintenance, and campaign sequencing through sequence-dependent cleaning times derived from site rules. Three objective formulations were examined on real industrial snapshots: makespan, makespan plus total tardiness, and makespan plus average tardiness.

Computational studies on instances with 10, 30, and 84 jobs showed consistent improvements over reference plans created on the same data snapshots. The proposed schedules achieve markedly shorter makespans, even though they respect a stricter feasible region that includes cleaning and calendar constraints. Tardiness measures also decrease with size. The composite objectives reduce late job counts with essentially unchanged makespan on the smaller instances and with a modest increase on the largest instance, which reflects the tighter capacity and the additional trade-offs implied by campaign sequencing. Optimality is certified for the small case, and medium and large cases terminate with small to moderate gaps under a fixed time limit, which is consistent with the growth in the number of campaign ordering choices.

The main practical insight is that a compact constraint programming model, fed by production data and site rules, can produce feasible and transparent schedules that improve both plan compactness and due-date adherence. The approach remains interpretable, as all scheduling elements are linked to plant calendars, routings, and cleaning tables.

There are several avenues for future work. First, machine learning can be integrated with the current pipeline to support enhanced data-driven clustering and decomposition [18, 19], for example by grouping products or sequences with similar cleaning behavior or by learning effective campaign blocks, which can reduce model size and speed convergence. Second, problem-specific heuristic components can be developed using adaptive efficient metaheuristic optimization frameworks [20, 21, 22].